\documentclass[12pt]{article}

\usepackage{graphicx}

\title{Analysis of the exotic resonances in the systems $QQ\bar q\bar q$ and $Qq\bar Q\bar q$ with the extended recoupling model}

\author{ Yu. A. Simonov \\

National Research Center ``Kurchatov Institute'' \\

Moscow, Russia}

\newcommand{\beq}{\begin{eqnarray}}

 \newcommand{\eeq}{\end{eqnarray}}

\newcommand{\be}{\begin{equation}}

 \newcommand{\ee}{\end{equation}}

\def\fun#1#2{\lower3.6pt\vbox{\baselineskip0pt\lineskip.9pt

\ialign{$\mathsurround=0pt#1\hfil ##\hfil$\crcr#2\crcr\sim\crcr}}}

\newcommand{{\SD}}{\rm SD}

\newcommand{{\Mc}}{\mathcal{M}}

\newcommand{\vex}{\mbox{\boldmath${\rm x}$}}

\newcommand{\vep}{\mbox{\boldmath${\rm p}$}}

\newcommand{\veq}{\mbox{\boldmath${\rm q}$}}

\newcommand{\veR}{\mbox{\boldmath${\rm R}$}}

\begin{document}

\maketitle

\begin{abstract}

The phenomenon of narrow peak $X(3875)$, discovered recently by the LHCb in the $D^*D$ system, and the absence of  resonances in the $D_sD_s$ and $D_s^*D_s^*$ systems is discussed within the new extended version of the Recoupling Model, where the resonance is the result of infinite recoupling transitions of confining strings  between quarks and antiquarks.
We show that existence of the resonance in the $D^*D$ and its absence in $D_sD_s$ and $D_s^*D_s^*$ systems are the natural results of the Recoupling Mechanism. The application of this mechanism to the $D\bar D^*$ system allows to obtain the resonance $Z_c(3900)$ with a larger width in agreement with experimental data.

\end{abstract}

\section{Introduction}

Recently the LHCb has published surprising data on the resonance  in the $D^0D^0\pi^+$ system \cite{1,2}, displaying a very narrow resonance $X(3875)$ near the $D^*D$ threshold with the mass $M= 3875$ MeV and the width below 1 MeV. The existence of so narrow resonance in the system of two strongly interacting hadrons is nontrivial and creates several problems for the theory:

\begin{itemize}
\item{\bf The problem 1}. {The resonance $X(3875)$ has the width $410$ keV \cite{1,2} which is much lower than in other resonances with the similar structure}. This narrow resonance in the $D^*D$ system seems to have dynamics  different from that in the $D^*\bar D$ system, where the resonances have much larger width \cite{3,4}, however in both cases the resonances appear  close-by the thresholds.

\item{\bf The problem 2}. {The resonances in the $D\bar D^*$  \cite{3,4}, $D_s\bar D^*$ \cite{5} systems have much larger widths}. Even rather narrow resonance $Z_{cs}(3985)$  in the $D_s^-D^{*0}$ system,  observed   nearby the thresholds \cite{5}, has the width around 12 MeV. Then the question arises whether internal dynamics of these resonances is universal, i.e. the same in $DD$ and $D\bar D$ systems.

Finally, \item{\bf The problem 3}. { The systems $D_sD_s$ and $D_s^*D_s^*$ similar to $DD^*$ do not show any resonances near the threshold region \cite{6}}. At this point one needs to find the mechanisms ensuring the resonance characteristics, both in the $D^*D$ and $D^*\bar D$ systems,
and precluding the existence of the resonances in the same region in the $DD,D^*D^*$ systems.
\end{itemize}

The purpose of present paper is to formulate a general scheme and suggest a concrete mechanism, which solves the problems stated above.

In our paper we will formulate the extended version of the Recouplimg Model (RM), suggested in original version in \cite{7}, to explain the resonances with similar properties $X(3875)$ and $Z_c(3900)$ and solve the problems 1,2,3. In particular, within this approach  we also show  that  in  the systems $DD, D^*D^*$ with two identical mesons the resonances do not appear.

In the literature  exotic four-quark resonances have been considered using different mechanisms, including standard coupled channel method \cite{8}, the tetraquark model within the Born--Oppenheimer approach \cite{9,10,11,12}, the cusp phenomenon and the triangle singularity \cite{13,14,15}, the few-body theory of hadron-hadron interaction \cite{16}, analytic approach including the boson exchange mechanism \cite{17,18,19,20}, and the QCD sum rule method \cite{21}.   In the present paper we formulate the model of the hadron-hadron interaction with the resonance creation as an extended version of the recoupling approach \cite{7},  which explains  the hadron-hadron resonances as the poles from the infinite sum of the recoupling diagrams, where the hadrons pass through the stage of the compound bag, producing  the string recoupling  between hadrons. In this version of RM no additional interaction between quarks or hadrons is introduced but as in \cite{7} only the exchange of the strings between two hadrons. As an example the hadron $h_1(Q_1,\bar q_2)$ meets the hadron $h_2(q_3,\bar Q_4)$  and it is assumed that the only visible transition is the exchange  of the confining string  between two hadrons --  forming two new hadrons - $h_3(\bar q_2,q_3)$ and $h_4(Q_1,\bar Q_4)$ , called the recoupling process. Notice that the nonsymmetric transition in the string recoupling takes place in the $D_1\bar D_2$ case, for example $D\bar D\to J/\psi \omega$, and this process with the probability amplitude $V_{12}$ is followed by the inverse transition with the amplitude $V_{21}$ and so on. In addition in the extended RM
we below introduce the intermediate stage of the compound bag containing all four (anti)quarks which allows to factorize
the amplitude $V_{ik}$ as $V_{ik}= V_i V_k$.

For two heavy-light mesons the string recoupling, e.g. in the $DD^*$ and $D\bar D^*$ processes, is quite different:
e.g. the $D^+D^{*0} \to D^{+*}D^0$ process is symmetric with almost coinciding thresholds, while $\bar D D^{0*}$ recouples into
the $J/\psi h$ system, where $h=\rho, \pi,..$ with different thresholds and dynamics. This can explain the problem 2. To understand  difference of the string recoupling situation in the $DD^*$ type of processes from the fully symmetric case of $D_s^+D_s^+$ one can compare the first process (with e.g. $\bar q_2$ different from $\bar q_4$) in the recoupling transition:
$h_1(Q_1,\bar q_2)$ plus $h_2(Q_3,\bar q_4)$ into $h_3(Q_1,\bar q_4)$ plus $h_4(Q_3,\bar q_2)$, with that of  identical hadrons ($Q_1=Q_3,~\bar q_2=\bar q_4$), where the total wave function is fully symmetric, so that the exchange of the quarks has no sense and the recoupling transition does not occur. This explains the absence of experimental resonances in the  $D_sD_s$ and $D_s^*D_s^*$ systems \cite{6}, thus resolving the problem 3. One can expect that in the $D_sD_s^*$ system, where the recoupling transitions are not forbidden, a  narrow resonance, similar to the case
of $DD^*$ -- the $X(3875)$ \cite{1,2}, could  exist.

Here we use the RM to  study the experimental process, where (among other products) specifically two hadrons (the pair 1) are produced,  which can transform into another pair of hadrons
(the pair 2) with the probability amplitude $V_{12}(\vep_1,\vep_2)$ and the relative momenta  $\vep_1,\vep_2$. We consider an infinite set of the transformations, where hadrons (between the acts of transformation) propagate freely with the Green's functions $G_i(E,\vep_i)~(i=1,2)$. Then the infinite set of transformations from the pair 1 to the pair 2, can be written as \cite{7},
$$
W_{12}= G_1 V_{12} G_2 + G_1 V_{12} G_2 V_{21} G_1 V_{12}G_2 + G_1 V_{12} G_2 V{21} G_1 V_{12} G_2 V_{21} G_1 V_{12} G_2 + ...=$$\be
= G_1 V_{12} G_2 (1 + N + N^2 + ...) = G_1 V_{12} G_2 \frac{1}{1- N},~~  N(E)= G_2 V_{21} G_1 V_{12}.
\label{eq.1}
\ee

As a result the total production amplitude $A_2$ of the pair 2 (in  possible complex with other particles) can be written  as a product of the slowly varying function $F(E)$ and  possible singular factor $f_{12}(E)= \frac{1}{1-N(E)}$, namely, $A_2= F(E) f_{12}(E)$ \cite{7}. In the RM \cite{7}  the resulting amplitude $f_{12}(E)$ was derived in the form,

\be
f_{12}(E)= \frac{1}{1- \frac{z}{(\nu_1-ik_1(E))(\nu_2-ik_2(E))}}.
\label{eq.2}
\ee
Here $k_i (i=1,2)$ is the relative momentum of two hadrons in the initial $i=1$ or the final state $i=2$. In the nonrelativistic limit for the energy values $E$ nearby the thresholds one has
\be
 k_i= \sqrt{2\mu_i(E-E_{thi})},~~ \mu_i= \frac{m_{ai}m_{bi}}{m_{ai}+m_{bi}}.
\label{eq.3}
\ee
From (\ref{eq.2}) it is clear what basic properties of the system are needed to display a sharp singularity: first,  the close distance
between the thresholds and second, the close values of the parameters $\nu_i,~i=1,2$. Those thresholds $E_{thi}~(i=1,2)$ are almost equal in the case of the $X(3875)$ \cite{1,2}; notice that the thresholds are also nearby each other for other resonances: $Z_c(3900), X(3915), Z_{cs}(3985)$. Here one should take into account that in the first approximation of the RM the parameter $z$ in (\ref{eq.2})  is considered as a constant, playing the role a fitting parameter, which is usually taken  around the value of the product $\nu_1\nu_2$ to sharpen the transition.

In what follows we shall use the extended version
of the RM formalism, where the initial hadrons in the stage 1 transform into the compound bag \cite{22} with the wave function $\Phi(q_i)$, containing four constituents
(quarks and antiquarks), which then decays into the final pair of hadrons. Namely,
\be
V_{12}=(\Psi_1(h_{a1}h_{b1})\Phi(q_i))|(\Phi(q_i)\Psi_2(h_{a2}h_{b2})= V_1(\vep_1)V_2(\vep_2).
\label{eq.4}
\ee

This factorization is essential in the extended RM and ensures  appearance of the narrow X(3875) due to the resulting symmetric
form, while the original non-extended direct recoupling form (as in \cite{7}) may violate this symmetry.
Nevertheless, the extended RM keeps the main feature of the recoupling process: no new quarks or antiquarks
participate in the process and therefore the symmetric form with $V_1= V_2$ is possible for the $DD^*$ system, while it is not possible for the $D\bar D$ or $D\bar D^*$ systems with the string recoupling, which necessarily impose the transfer into $J/\psi \rho$ or $J/\psi \omega$. However the latter thresholds are also rather close to those of $D\bar D^*$ and more wide resonances are possible.

The explicit form of $N(E)$ in \ref{eq.1} and resulting parameters of the extended RM are given in next section.

\section{The Extended Recoupling Model}

As one can see in (\ref{eq.2}), the main problem now is to calculate the parameters $\nu_1,\nu_2$ (beyond the parameter $z$ which we
shall use as a free parameter at the present stage) and this will be done below. In this section we will give the details of the extended RM, where the recoupling amplitude $V_{12}$ (\ref{eq.4}) factorizes into two factors $V_1(\vep_1)$ and $V_2(\vep_2)$, which have the following explicit form, e.g. for $V_1$,

\be
V_1(\vep_1)=(\Psi_1(h_{a1}h_{b1})\Phi(q_{i})),
\label{eq.5}
\ee
where in (\ref{eq.5}) the integral enters over all four quark momenta, while the first factor is the hadron-hadron wave function and the second one is the compound bag wave function \cite{22}, namely,

\be
\Psi_1(h_{a1}h_{b1})= \exp{i\vep_1(\veR_{a1}(x_1,x_2)- \veR_{a2}(x_3,x_4))}\psi_{a1}(x_1- x_2)\psi_{a2}(x_3-x_4).
\label{eq.6}
\ee
Here the c.m. coordinates of the mesons are defined via the effective quark energies $\omega_i~(i=1,2,3,4)$ (see Appendix 2 and \cite {23}) inside the hadrons,  namely, $\omega_1,\omega_2$ in the hadron $h_{a1}$ and $\omega_3,\omega_4$ in the hadron $h_{a2}$, and in (\ref{eq.7}) one can write $R_{a1}=b_1 x_1 + b_2 x_2,~ R_{a2} =b_3 x_3 + b_4 x_4$, where $b_1=\frac{\omega_1}{\omega_1 + \omega_2}, ~i= 1,2$ and
$b_i= \frac{\omega_i}{\omega_3 + \omega_4}, i= 3,4$.

The compound bag wave function $\Phi(x_1,x_2,x_3,x_4)$ is the  product of all four quark (antiquark) wave functions in the compound linear confining potential, created by the confining strings \cite{22}. It is convenient to use the equivalent Gaussian form of the wave functions, namely

\be
\Phi({x})= \prod {i} \exp\{-\lambda_i(\vex_i- \veR)^2\}, \veR= \sum_i \beta_i \vex_i, \beta_i= \frac{\omega_i}{\sum \omega_i}.
\label{eq.7}
\ee

As was said above, the variables $\omega_i$ are the effective quark energies in the hadrons, which are well defined in the relativistic $q\bar q$ formalism \cite{23} (they are given  in the Appendix A1 of \cite{7}), while in the Gaussian form of the confined (compound) wave function the variable  $\lambda_i$ is expressed via the string tension $\sigma= 0.18$~ GeV$^2$ and the $\omega_i$,
namely, $\lambda_i=\frac{\sigma \omega_i}{2.27}$. Thus both $\beta_i,\lambda_i$ are defined by the quark energies $\omega_i$
and one must introduce the final parameters of the hadron wave functions for
the hadrons $h_{a1},h_{a2}$, which we choose also in the Gaussian form, well describing the light and charmed meson wave functions in the
intermediate region \cite{24}: $\psi_{ai}(\veq)= {\rm const} \exp\left(-\frac{\veq^2}{2\delta_i^2}\right)$. In this way each of the terms $V_1,V_2$ in
(\ref{eq.4}) is defined by the set of four effective quark energies ${\omega_i}$ and  two hadron wave function parameters $\delta_1,\delta_2$ which are given  in the Appendix A3 in \cite{7}.

The resulting form of the matrix element $V_1(\vep_1)$ and the parameter $\nu_1$ can be derived in the following form,

\be
V_1(\vep_1)= {\rm const} \exp(-Y_1 \vep_1^2),~~ \nu_1 \approx \frac{1}{\sqrt{2Y_1}}.
\label{eq.8}
\ee
In a similar way one obtains $V_2(\vep_2)$ and $\nu_2$ where all parameters $\omega_i,\beta_i,\lambda_i,\delta_i$
can be different. Explicit values of these parameters and the resulting values of $\nu$ for the systems $DD^*$ and $D\bar D^*$ (the case of
$X(3875)$ and $ Z_c(3900))$  are given in the Appendix A1 of this paper and discussed in the next section.

\section{Application of the extended RM to the resonances $X(3875)$ and $Z_c(3900)$}.

{\bf The example 1}
{\bf $X(3875)$}\\

The first and probably the most striking effect, produced by the resonance $X(3875)$, was observed by the LHCb group \cite{1,2}
in the system $D^{*+}D^0,D^{*0}D^+$ with the properties:   $X(3875),~J^P=1^+,~\Gamma = 410$ ~ keV, \cite{1,2}, for which $[D^{*+} D^0 \to D^{*0}D^+],~ E_1= 3.875$~GeV,~
$E_2= 3.876$~GeV. The reduced masses have very close values: $\mu_1= \mu_2= 0.967$~GeV. In the RM this resonance can be created  due to infinite chain of the recoupling
transitions: $D^*(c\bar d)D(c\bar u) \to D^*(c\bar u)D(c\bar d)$, where light antiquarks $\bar d,\bar u$ always change their hosts (mesons).

In this case one has two close-by thresholds: $E_{th1}= M(D^{*+}) + M(D^0)= (2010 + 1865)$~MeV= 3875~MeV and $E_{th2}= M(D^{*0} + M(D^+)=
 (2007 + 1869.6)$~MeV = 3876.6~MeV. Our purpose now is to find $\nu_1,\nu_2$
for both stages of the reaction. Using the effective quark energies in the $D,D^*$ mesons from the Appendix A2:  $\omega_i= 1.4,0.47,1.51,0.5$ (all in GeV),
and the parameters of the meson wave functions: $\delta_1= 0.48$~GeV,~$\delta_2= 0.49$~GeV from Appendix A3 of \cite{7}, one can calculate $\nu_i$ in both stages
of the transitions with the use of the equation  (\ref{A1.1}). The value of the parameter $Y$ (\ref{eq.8}) can be obtained from the (\ref{A1.1})-(\ref{A1.4}) in the Appendix A1,
with the result $Y= 2.36$~GeV$^{-2}$.Then one obtains both $\nu_i$, coinciding within 1\%  accuracy,
$\nu_1= \nu_2 \approx \frac{1}{\sqrt{2Y}}= 0.46$~GeV.

The resulting singularity in the amplitude $f_{12}(E)$ can be found in the form of  (\ref{eq.1}) and for equal threshold masses it produces a pole nearby thresholds. Note that below in the Table~
\ref{tab.01} we increase the distance between thresholds up to $2$ MeV to make the cross section structure more visible.

The actual singularity structure is more complicated and consists of four poles and two square-root
branch points, which possibly contribute to the width $\sim 0.4$ MeV of the resonance, as seen from experiment \cite{1,2}.
To define the structure of experimental cross sections, firstly, we take the recoupling parameter $z=0.2$~GeV$^2$ and putting the parameters in (\ref{eq.1}),  we obtain the distribution $|f_{12}(E)|^2$  with the maximum at $E= 3876$ MeV and the values of $|f_{12}(E)|^2$, given below in the Table  \ref{tab.01}.

\begin{table}[h!]

\caption{The values of the $|f_{12}(E)|^2$ near the channel thresholds for the transition {\bf 1)} }

\begin{center}

\label{tab.01} \begin{tabular} {|c|c|c|c|c|c|c|} \hline

$E$~(GeV)& 3.862& 3.872& 3.874& 3.876& 3.878&3.884\\

$|f_{12}(E)|^2$& 4.39&11.9&35.6& 48.9& 10& 3.67\\

\hline

\end{tabular}

\end{center}

\end{table}

As one can see in the Table~\ref{tab.01},  the position of the resulting resonance is just at $E= 3.876$ GeV and the
width $\Gamma\approx 2$ MeV, which is close to the assumed distance between thresholds; the width can be derived smaller by adjusting the value of $z$.
Indeed, for $z= 0.25$ GeV$^2$ one obtains  much larger value of
$|f(E)|^2$, exactly at the lower threshold, $E= E_1= 3.874$ GeV with the width around 1 MeV. One can see that the varying of the parameter $z$ does not change much  resulting position
of the maximum in  $|f(E)|^2$
but can increase the resulting width of the resonance. Thus this example shows that in this case in the extended  RM a good agreement with experimental data is obtained.\\

{\bf Example 2}.
{\bf  The resonance $Z_c(3900)$ in the $\bar D^*D$ system \cite{3,4}}.\\

This resonance, named $Z_c(3900)$, was observed  by the BESIII in the reaction $e^+e^- \to \pi D\bar D^*$ \cite{3} and $e^+e^- \to
\pi^+\pi^- J/\psi$ \cite{4}.

Here we assume that the resonance in this system appears as due to infinite set of recoupling transitions $D^+\bar D^{-*} \to J/\psi \omega$. We shall consider  the string recoupling  in the framework of the extended RM. As the first step one defines the parameters $\nu_1,\nu_2$ in the
basic equation (\ref{eq.2}), using the formalism, given in  Appendix A1,  with the parameters of  these  transitions taken from
Appendix A2 for the hadrons $J/\psi,\omega,D,D^*$, where $\omega$ can be replaced by $\rho$. As a result the basic parameters of the step 1 - $D\bar D^*$ coincide  (with the accuracy of calculations) better than one percent, with those of Example 1:

$Y_1= 2.42$~GeV$^{-2}$ and $\nu_1\approx \frac{1}{\sqrt(2Y_1)} = 0.46$~ GeV.

However, for the step 2 - $J/\psi \omega$  from Appendix A1 and A2 one obtains the smaller parameter $\nu_2= 0.21$~GeV.
This fact of different values of $\nu_1,\nu_2$ in connection with a large difference in threshold energies might cause a wide
or no resonance at all, however in this case the thresholds appear quite close to each other. Indeed,the thresholds for the transition $J/\psi \rho^+ \to \bar D^0 D^{*+}$ are $E_1= 3875.14, E_2= 3879$~MeV. For the neutral channel $\bar D^-D^{*+}$ both thresholds in channels 1 and 2 coincide at $E= 3879$~MeV, and we below consider less singular situation
of noncoinciding thresholds. Here parameters $\mu_1= 0.967,\mu_2= 0.623$~GeV. Choosing the parameter $z= 0.1$~GeV$^2$ in (\ref{eq.2}), one obtains the  factor $|f_{12}(E)|^2$  given  below, in the Table  ~\ref{tab.02}.

\begin{table}[h!]

\caption{The values of the transition probability as a function of energy in the transition {\bf 3)} }

\begin{center}

\label{tab.02} \begin{tabular} {|c|c|c|c|c|c|c|} \hline

$E$(GeV)& 3.872& 3.875& 3.877& 3.879& 3.882& 3.885\\

$|f_{12}|^2)$& 6.64& 19.76& 21.88& 27.6& 3.96& 2.34\\

\hline

\end{tabular}

\end{center}

\end{table}

The numbers from the Table 2 show the sharp peak at $E= 3879$ MeV and the width  $\sim 10$~ MeV. This result approximately
agrees with the experimental position of the resonance and its experimental width,  $\Gamma(Z_c(3900)) \approx 28$~MeV \cite{3,4}.
If one slightly varies the fitting parameter $z$, then the position of the resonance remains near the thresholds while the width can strongly increase for large values $|z-\nu^2|$.

\section{Conclusions}

In our paper the basic problems of the $X(3875)$ and $Z_{c}(3900)$  resonances have been formulated as the  problem 1 -- why  very narrow resonances exist in the four quark  system $cc\bar q\bar q$,and the problem 2- why similar but somewhat wider resonances exist in the $c\bar c q\bar q$ system with seemingly different dynamics, and finally -- the absence of the resonances in the systems with identical mesons, like $DD$, was put as the problem 3.

It is not easy to find the dynamics which can solve simultaneously all three problems. In our  paper we have shown that the extended RM allows to find the way to solve all three problems. In the case of in the $X(3875)$ resonance the very narrow resonance is created  due to the infinite process  of the string recoupling in the $DD^*$ system, $D^+D^{0*} \to D^{+*}D^0$, with  almost identical thresholds and the dynamics in channels 1 and 2. This allows to solve the problem 1.

It is interesting that in the $D\bar D, D\bar D^*$ systems the string recoupling necessarily leads to the completely different transition channel 2,  e.g. to $J/\psi\omega$ with different thresholds and dynamics, and due to that  with larger width. This allows to solve the problem 2. In the same way one can analyze the observed resonances $Z_c(3900),X(3960),X(4140),$ $Z_{cs}(4000)$.

Finally, the extended RM does not predict resonances in the fully symmetric systems with identical pairs, like $DD$ and $D^*D^*$, thus solving  the problem 3. One can envisage many other  applications of the extended RM to different hadron-hadron systems,  with possible inclusion of many-channel transitions possible in the $D\bar D$ systems.

The author is grateful to A. M. Badalian for very useful discussions and to N. P. Igumnova for collaboration.

\section*{\bf Appendix A1. Calculation of the parameters $\nu_1,\nu_2$ for the $X(3875), Z_c(3900)$ systems}.

 \setcounter{equation}{0} \def\theequation{A1.\arabic{equation}}

As it was shown in first two sections, the matrix element $V_i(\vep_i)$ is defined by the integral of the product of two
wave functions, e.g.  $V_1(\vep_1)=(\Psi_1(h_{a1}h_{b1})\Phi(q_{i}))$, where the parameters  $\omega_i,~i=1,2,3,4, ~\delta_1,\delta_2$, as well as the parameters $\beta_i,\lambda_i$, are expressed via the primary ones. Namely, $b_i$ are defined via $\omega_i$ in (\ref{eq.6}) and $\beta_i$ in (\ref{eq.7}).

As a result of the calculations the basic quantity $Y$ can be presented  in the following form,
\be
Y= A - \frac{B_1^2}{4f_1} - \frac{(B_2-\frac{B_1D}{2f_1})^2}{4(f_2-\frac{D}{4f_1})},
\label{A1.1}
\ee
where the following notation is used,

$$
f_1= C_1 + 1/(2\delta_1^2), f_2= C_2 + 1/(2\delta_2^2),
A= \sum_1^3 \epsilon_i a_{i}^2, B_k= 2\sum_1^3 \epsilon_i a_{i} d_{ki}, k= 1,2.$$
\be
C_k= \sum_1^3 \epsilon_i d_{ki}^2, k= 1,2; D= 2 \sum_1^3 \epsilon_i d_{1i}d_{2i},
\label{A1.2}
\ee
and $\epsilon_i$ are defined as

\be
\epsilon_i= 1/(4\lambda_i) - \frac{\beta_i}{4\beta_4 \lambda_i}\frac{1}{1/(4\lambda_i) + \sum_{k=1}^{k=3} \beta_k/(4\beta_4\lambda_k)},
\label{A1.3}
\ee
where $a_i,d_{ki}$ are expressed via the sets ${b_i},{\beta_i}$ of the participating (anti)quarks,

$$
a_1= -(b_1 + b_4 (\beta_1/\beta_4)),~~ a_2= -(b_2 + b_4(\beta_2/\beta_4)),~~ a_3= b_3- b_4 (\beta_3/\beta_4),$$\be
d_{11}= -1,~~d_{12}= 1~~, d_{13}= 0,~~ d_{21} = -\beta_1/\beta_4,~~ d_{22}= -\beta_2/\beta_4,~~ d_{23}= -1 -\beta_3/\beta_4.
\label{A1.4}
\ee

\section*{\bf Appendix A2. The definition of the effective quark energies in the mesons}.

 \setcounter{equation}{0} \def\theequation{A2.\arabic{equation}}

Following the relativistic theory of hadrons with the confinement interaction \cite{23}, one can write the Hamiltonian in the c.m. system of two quarks in the form

\be
H= \frac{m_1^2}{2\omega_1}+ \frac{m_2^2}{2\omega_2}+ \frac{\omega_1+ \omega_2}{2}+ \frac{\vep^2}{2\tilde\omega}+ \sigma r + V_g
\label{A2.1}
\ee

Here $\tilde{\omega}= \frac{\omega_1 \omega_2}{\omega_1+\omega_2}$ and $V_g$ implies the gluon-exchange interaction, which here is neglected for simplicity.

Solving the equation $H \psi= M \psi$, one obtains the expression for the spin-averaged hadron mass of the same form, as in (\ref{eq.A2.1}), where instead of three last terms the energy eigenvalue
enters:  $\epsilon_n(\tilde{\omega})= (2\tilde{\omega})^{-1/3} \sigma^{2/3}a(n)$

\be
M_n= \frac{m_1^2}{2\omega_1}+ \frac{m_2^2}{2\omega_2}+ \frac{\omega_1+ \omega_2}{2}+ \epsilon_n(\tilde{\omega}),
\label{eq.A2.2}
\ee
where $a(n)$ is the Airy number and  $a_0=2.338$. In our approach \cite{23}, which has allowed to calculate hundreds of the lowest hadrons \cite{24}, the values of $\omega_i$ are obtained from the extremum condition:

$\frac{\partial M(\omega_1,\omega_2)}{\partial \omega_i}= 0$, which allows to find the extremum values $\omega_i^{0}$ . As a result, the extremum  values of $\omega_1,\omega_2$ for the heavy-light mesons $D,D_s,D^*,D_s^*$ are approximately related as 3 to 1, producing the coefficients e.g. $b_1=0.75,b_2=0.25$. In this way one can calculate all other coefficients, participating in the formulas of Appendix A1.

\end{document}